# 2D Mn Doped MoS$_2$: An Efficient Electrocatalyst for Hydrogen Evolution Reaction

Joy Ekka[1], Shrish Nath Upadhyay[2], Verma Bunty Sardar[1] and Srimanta Pakhira[1, 2, 3*]

[1] Discipline of Physics, Indian Institute of Technology Indore (IITI), Simrol, Khandwa Road, Indore-453552, MP, India.
[2] Discipline of Metallurgy Engineering and Materials Science (MEMS), Indian Institute of Technology Indore, Indore-453552, MP, India.
[3] Centre for Advanced Electronics (CAE), Indian Institute of Technology Indore, Indore-453552, MP, India.
*Corresponding author: spakhira@iiti.ac.in (or) spakhirafsu@gmail.com

## Abstract:

Earth-abundant two-dimensional (2D) pristine transition metal dichalcogenides (TMDs) have emerged as a superlative class of materials for several applications in electronic devices, energy storage devices, gas sensing, etc., and they have recently attracted great attention, owing to their good catalytic activity and excellent stability toward electrochemical H$_2$ Evolution Reaction (HER). Each individual layer of the TMDs consists of three atomic layers in which the transition metal is sandwiched by two chalcogens. Furthermore, the chalcogen atoms are saturated and therefore they are not highly reactive. Because of its robustness, 2D monolayer MoS$_2$ TMD is the most studied material in this family and especially for the applications of electrocatalysis and H$_2$ evolution reactions (HER). Most of the experimental and theoretical studies have shown that the basal planes of the 2D monolayer 2H-MoS$_2$ are catalytically inert and hence, they cannot be used directly in desired applications such as electrochemical HER. To activate the inert basal plane of the pristine 2D TMDs, it is needed to create some defects or doping of some heteroatoms in the pristine TMDs. Phase engineering techniques have been used to activate the basal plane of the 2D TMDs. In this article, we have computationally developed 2D monolayer Mn-MoS$_2$ material and its application in HER. Stable S terminated edge of the MoS$_2$ shows low catalytic activity due to its inert basal plane, so to exploit these edges for improved performance we doped Mn in the pristine MoS$_2$ material. Using trailblazing and state of the art first principles-based density functional theory we performed methodical and rigorous inspection of electronic structures and properties of monolayer Mn doped $MoS_2$ to be a promising alternative to noble metal-based catalysts for HER. Periodic 2D slab of $Mn-MoS_2$ was created to study the electronic properties and the reaction pathway occurring on the surface of the material has been delved into by creating $Mn_1Mo_9S_{21}$ molecular cluster model. Our study reveals that the 2D $Mn-MoS_2$ monolayer based catalyst follows the Volmer-Heyrovsky reaction with very low energy barriers during the HER mechanism. This study is focused on designing a low cost and efficient electrocatalyst for HER by using earth abundant TMDs and lowering the activation barriers by scrutinizing the kinetics of the reaction for reactivity.

# 1 Introduction:

World's energy supply is heavily dependent on fossils fuels and energy production emanating from this source releases substances that are harmful for the environment. Rapid depletion of this source is also another concern which drives the necessity to find eco-friendly and zero emission energy sources. To resolve this, we need to increase the usage of renewable energy sources to meet the future energy demands. Hydrogen offers itself as a new and unconventional renewable wellspring due to its high energy density and zero emission greenhouse gases among other available fuels.[1] To produce clean and efficient hydrogen, electrolysis of water has been a viable way ever since its discovery in 1789.[2] The electrochemical water splitting is a sustainable strategy to produce hydrogen (and oxygen) and replace conventional fossil fuels. Hydrogen is a non-polluting energy source as water itself is a byproduct of $H_2$ combustion. However, the conventional water splitting process cannot just separate water into hydrogen and oxygen, and just like any other chemical processes even this reaction needs energy input to overcome the barrier in the electrochemical process. Thus, the electrochemical water splitting requires highly active catalysts to bring down the overpotential needed to produce hydrogen. HER i.e., simply $2H^+ + 2e^- \rightarrow H_2$ is a multiphase reaction and can occur via two possible processes, either a Volmer-Heyrovsky process or Volmer-Tafel process.

However, the mentioned reactions are competitive process, and they are dependent on the electronic structure of the electrode surface. Boosting the efficiency of sluggish HER process is a challenging task at hand. As of now Platinum (Pt) is the best known catalyst for HER due to its zero overpotential in acidic electrolytes.[3] Because of the optimum Gibbs free energy for adsorption of atomic hydrogen, hydrogen binding energy and desorption of hydrogen from the surface with low activation energies, Pt based catalyst have been long known as effective HER catalysts. However, the limited availability and high cost of Platinum inhibit its usage as an efficient catalyst at large scale in commercial applications and industries. To produce hydrogen at a global scale we need to cut down the cost of production. Reducing the dependency on noble metal-based catalysts or to reinstate them even completely with non-noble metal alternatives would be a driving step towards hydrogen economy. So, the urge for search of Pt free catalyst for HER is of paramount importance in modern materials science and technology.

Countless exhaustive research works have been focused on materials which are excellent electrocatalysts for HER.[4–9] Plethora of breakthroughs has been achieved in this regard for rational design of HER electrocatalysts. In recent times, earth abundant layered transition metal dichalcogenides (TMDs) for example $MoS_2$ has attracted tremendous interest due its inherent properties which produce hydrogen at very low overpotential with high current density.[10] This greatly recognized that the 2D monolayer $MoS_2$ material can exist in several possible structures such as 1T (octahedral structure), 1T´ (distorted octahedral structure) and 2H (hexagonal structure).[11] The 2H phase structure (i.e. 2D monolayer 2H-$MoS_2$) is the most stable[11] and they have been commonly used for HER electrocatalyst.[12–14] The high activity of $MoS_2$ is because of its appropriate Gibbs free energy of adsorbed atomic hydrogen.[15] Although this material shows promising aspects but it is still not sufficient in its present form for large industrial and commercial application purposes due to its inert basal plane. Theoretical calculations indicated that the P-doped $MoS_2$ shows a good catalytic activity for HER by reducing the change of free energy ($\Delta G$). This is due to the P-doping in the pristine MoS2 which activated the inert basal plane of it,[16] but, this non-metal doping is quite difficult because of instantaneous formation of MoP.[5] To avoid formation of MoP, quite expensive apparatus such as plasma ion implantation are required and in some cases especial precautions has to be taken during the P-doping in the MoS2.[17] Thus, another method or material is required for developing low cost and efficient HER catalysts to produce desirable $H_2$ for industrial applications. Another problem frequently appears in the 2D TMDs due to the stacking feature of the 2D layers that decreases the amount of exposed sites and the conductivity along two stacked layers is extremely low,[18] thereby impeding charge transfer and decreasing the HER performance of the 2D TMDs. A prominent factor controlling the rate of HER is vested upon the fact that the pristine 2D $MoS_2$ shows semiconducting properties indicating low conductivity for electrons, thus being inadequate for large commercial applications. One of the promising ways for enhancing HER is to expose the active site of the pristine 2D monolayer 2H-MoS$_2$.[19] It was found that the most of the active sites of the pristine 2D monolayer 2H-MoS$_2$ for HER are located at the Mo and S edge sites.[20] In order to modulate the electron transport for achieving proper conducting pathway and enhance the hydrogen evolution, the doping of external elements in the pristine 2D monolayer 2H-MoS$_2$ nanostructure appropriately is the promising way in the modern technology.[9] Therefore, the mechanistic insights are relevantly important while designing efficient electrocatalysts for H$_2$ evolution.

The development of operative, stable, and economic HER catalyst to overcome the challenges associated with H$_2$ production from water electrolysis is a salient comprehension for driving down the production cost and extension of hydrogen economy. We hypothesize that Mn- doping in the 2D monolayer MoS$_2$ can activate the inert basal plane and Mn-doped 2D MoS2 can be a promising materials for an efficient H$_2$ evolution. For the transition metal based catalysts, their performance is correlated to their surface electronic structure and the electronic configuration of the *d*-orbital of the transition metal.[6] In this regard, we have computationally developed two dimensional (2D) single layer Mn doped MoS$_2$ material (in short Mn-MoS2) and investigated their electrocatalytic performance for efficient HER. First, we performed first principles-based hybrid periodic density functional theory (DFT)[21–26] calculations to obtain electronic properties like electronic band structure, band gap and density of states (DOS). We found out that 2D monolayer Mn-MoS$_2$ shows zero band gap due to Mn-doped in the pristine MoS2. The density of states calculations indicate that there is a large no. of available electronic states around the Fermi energy ($E_F$) level with a high availability of electrons due to the doping of Mn atoms in the pristine 2D monolayer MoS2 material.

One of the key features determining the smooth flow of reaction is the change in free energy (ΔG) of the possible reaction intermediates. So, to screen an appropriate candidate among the options available, it is important to compute the value of ΔG during hydrogen adsorption and this is an important parameter for evaluating the catalytic activity during HER process. Lately the quantum computation method has provided practicable procedures for calculating the free energy changes based on the density functional theory.[27,28] By modeling the possible reaction intermediates occurring during the hydrogen evolution process taking place on the surface of the electrocatalyst, thermodynamical properties can be obtained using DFT methods. Therefore, we prepared a molecular cluster model system of the Mn-MoS$_2$ material and carefully studied each and every reaction intermediate occurring during the HER by finite modeling based DFT based calculations in both gas and solvent phase. Our study showed that Mn-MoS$_2$ shows excellent catalytic activity.

# 2 Methodology and Computational Details:

## 2.1 Periodic Structure Calculations

We have systematically investigated the electronic properties calculations i.e., band structures and density of states (DOS) of both the pure MoS$_2$ and Mn doped MoS$_2$. The periodic structure computations and the equilibrium structures were obtained by performing hybrid dispersion corrected periodic density functional theory (in short DFT-D) (here B3LYP-D3 method)[29–38] implemented in *ab initio* CRYSTAL17 suite code.[39] The electronic properties calculations were obtained by using the same B3LYP-D3 method.[40–44] In the present calculations, we have accounted for the weak long-range van der Waals (vdW) effects[45] resulting from the interaction between atoms by including the semi empirical corrections (Grimme's–D3 corrections).[36] The weak vdW interaction between the layers of both the materials has been included in the present DFT calculations by adding Grimmes's semi-empirical dispersion parameters.[30–34] Triple-ζ valence with polarization function quality (TZVP) Gaussian basis sets were used for Sulphur (S)[46,47] and Manganese (Mn)[46] atoms, and HAYWSC-311 (d31) G type basis sets with Hay and Wadt small core effective pseudopotential for Molybdenum (Mo).[48] DFT-D method provides a good quality geometry of the 2D layered structure material after reducing the spin contamination effects such that it will not show any effect on the electronic structure and electronic properties calculations (i.e. band structure and the total density of states (DOS)).[23,33,49–52] The threshold used for evaluating the convergence of the energy, forces and electron density was set to 10$^{-7}$ a. u. for each parameter. The height of the unit cell was formally set to 500 Å (which considers there is no periodicity in the z-direction in the 2D slab model in CRYSTAL17 code), *i.e.* the vacuum region of approximately 500 Å was considered in the present calculations to accommodate the vacuum environment.[24,53] The unit cell of the 2D monolayer MoS$_2$ has been extended to a 3 x 3 x 1 to form a supercell and Mn atoms were doped by replacing the Mo atoms. It was found that the Mn-doping concentration was 12.5% in the 2D Mn-MoS2 material. In the atomic structure relaxation simulation, a vacuum slab of 500 Å was inserted between the layers to avert the interlayer interaction. For the periodic 2D layer structure (i.e. 2D slab) computations, a single 2D TMDs (here both the MoS2 and Mn-doped MoS2) layer terminated on the (10$\bar{1}$0) (Mo-/Mn-edge) and ($\bar{1}010$) (S-edge) boundaries with three Mo per unit cell has been considered as shown in Figure 2. It should be mentioned here that the exposed surfaces are generally the (001) basal plane of the S−Mo−S (Mn-doped in the case of Mn-MoS2) tri-layer, the Mo-/Mn-edge ($10\bar{1}0$) and S-edge ($\bar{1}010$).

The electronic band structures and total DOS calculations have been performed at the equilibrium structures of the TMDs by employing the same DFT-D method. All the integrations of the first Brillouin zone were sampled on 20x20x1 Monkhorst-pack,[54] k-mesh grids for the pristine 2D MoS$_2$ and 4x4x1 for 2D Mn-MoS$_2$ . The threshold for the convergence of energy, forces, and electron density was set to be $10^{-7}$ a.u. for evaluation. The k-vector path taken for plotting the band structure was selected as $\Gamma - M - K - \Gamma$ for both the materials (i.e., pristine MoS$_2$ and Mn-MoS$_2$). The atomic orbitals of Mo, S, and Mn were used to compute and plot the total DOSs. The single point calculation has been done from the non-normalized wave function at zero Kelvin temperature with respect to vacuum. To create the graphics and analysis of the crystal structures studied here a visualization software VESTA[55] was used.

## 2.2 Finite Cluster Modeling

Further we developed a cluster model system for both the 2D MoS$_2$ and Mn-MoS$_2$ materials of our material to investigate HER mechanism by using GAUSSIAN 16[28] suite code to study the reaction mechanism. A non-periodic finite molecular cluster model Mo$_{10}$S$_{21}$ system for the pristine MoS$_2$ and Mn$_1$Mo$_9$S$_{21}$ for the Mn-MoS$_2$ (as shown in Figure 1) has been considered here to investigate HER in both the gas phase and solvent phase calculations, and the M06-L[56,57] DFT method has been applied to investigate the reaction pathways, kinetics, barriers, and mechanism. Figure 1 shows how we extract a triangular cluster from the periodic array to expose only the Mo edges. Figure 1 shows the schematic representation of the finite cluster Mn$_1$Mo$_9$S$_{21}$. This M06-L DFT method is a technique used for energetics, equilibrium structures, thermochemistry, and frequency calculations of the molecular cluster structures, and it has been found that the M06-L DFT method provides reliable energy barriers for reaction mechanisms of organometallic catalysts.[4,5,25,56–58] We focused on the energy barriers and changes of free energy during reaction to explore the reaction pathways by employing Minnesota density functional based on the meta-GGA approximation which is intended to be good and fast for transition metals.[56,57,59] We used 6-31G** Gaussian basis set for H[60,61], S[62],

O[63] and Mn[64] atoms, while LANL2DZ with effective core potentials for Mo[65,66] atom. Different transition states (TSs) were computed at optimized geometry and to visualize them ChemCraft[67] was used. Moreover, the Heyrovsky reaction mechanism was studied by deliberately adding three water molecules and a hydronium ion in the vicinity of the intended reaction region. The water cluster model $4(H_2O+H^+)$ was prepared as follows:- 4 water molecules were placed adjacent to each other connected via hydrogen bond and a proton was attached to one of the water molecule. This model was prepared to simulate the reaction of $H_2$ formation during Heyrovsky process which has been depicted in Figure 5h.

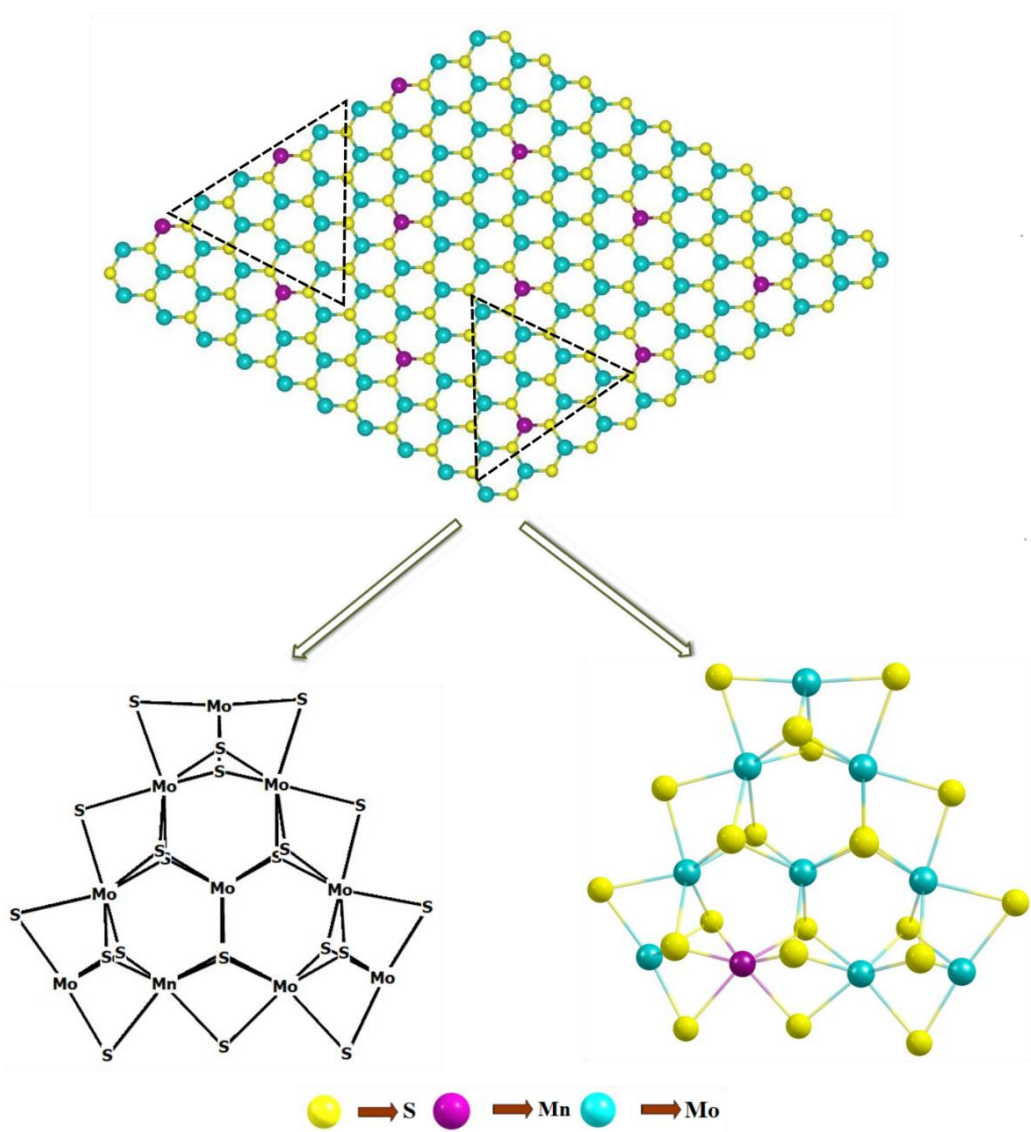

**Figure 1:** Cluster model ($Mn_1Mo_9S_{21}$) as derived from $Mn-MoS_2$ (represented by the dotted triangle).

To model the solvation effects we used polarizable continuum model[68] (PCM) with water as a solvent. The PCM method using GAUSSIAN 16 uses an external iteration method where the energy in the solution is computed by making the solvent reaction filed self-consistent with solute electrostatic potential, and as the real reaction is to take place in solvent phase. We modeled our reaction mechanism in water with static dielectric constant of 78.35. The geometric optimization and molecular energy in the solvent phase were calculated using the above-mentioned method.

# 3 Results and Discussions

## 3.1 Periodic vacuum slab inference

The relaxed lattice parameters and average bond distances are listed in Table 1. The lattice parameters (*a* and *b*) and the average Mo-S bond distance of the pristine $MoS_2$ 2D monolayer obtained by the DFT-D method are consistent with the previous reported results.[69] The values of lattice constants *(a = b) are* 3.18Å and the Mo-S bond distance is about 2.41Å which accord with the previous reported values, and it has a hexagonal 2D layer ***P6m₂*** symmetry.[69] This is a good estimation as compared to a work where the 2D monolayer $MoS_2$ structure was doped with 4% impurity and the bond distance was 2.39 Å between the Mn and the nearest S atom.[69] The doping of transition metal in the 3x3x1 supercell of the 2D monolayer $MoS_2$ has changed its symmetry from ***P6m₂*** to ***P1*** when the 2D Mn-$MoS_2$ has been formed. The average bond distance between the Mn and the nearest S atom was computed to be 2.30 Å which agrees well the previous result within 0.09 Å.[69] From the electronic properties calculations obtained by the same DFT-D method, we observed a direct band gap about 2.6 eV at *K* point in the Brillouin zone of the pristine 2D single layer $MoS_2$ material as shown in Figure 2a which is well harmonized with the previous theoretical and experimental results. The computed electronic band gap is slightly lower than the band gap obtained by GW approximation of monolayer $MoS_2$ which was 2.8 eV.[70] The Fermi level ($E_F$) was found at -6.36 eV depicted in the non-normalized band structure and DOS calculation as shown in Figure 2a highlighted by dotted blue color. After Mn-doping in the pristine 2D moS2 materials, the band structures have been changed and it was computationally found that the bands are overlapped around the $E_F$ as shown in Figure 2b. The present DFT-D study shows that the Fermi level was found at -5.04 eV in the case of the 2D monolayer Mn-MoS$_2$ with zero band gap indicating conducting character of the material. In other words, this zero-band gap suggests that the Mn-doping in the pristine TMD makes the 2D semi-conducting MoS$_2$ material to a conducting material in nature. This can also be justified by computing the electron density contribution from the 3*d*-subshell of the Mn atoms doped in the 2D monolayer MoS$_2$ material (as it can be seen from the *d* subshell DOSs in the right-hand side of Figure 2b). In other words, due to the addition of Mn atoms in the pristine MoS$_2$ to form the 2D monolayer Mn-MoS$_2$ material, the electronic band gap of it was decreased to zero depicted in the band structures and DOS calculations in Figure 2b. The addition of Mn to the pristine 2D TMD MoS$_2$ changes the electron accumulation in the bands as shown in the DOSs suggesting a high electron mobility with an indication of good catalytic activities for HER.

**Table 1: Lattice Parameters of the 2D monolayer MoS$_2$ and Mn-MoS$_2$ materials computed by the hybrid periodic DFT-D method has been provided here.**

| System | Lattice constants (a=b) | Interfacial angles (α, β and γ) | Space group symmetry | Average bond distance | |
|---|---|---|---|---|---|
| | | | | Mo-S | Mn-S |
| MoS$_2$ | 3.138 Å | α = β = 90.0° and γ = 120.0° | *P6m$_2$* | 2.411Å | --------- |
| Mn- MoS$_2$ (3x3 supercell) | 9.451 Å | α = β = 90.0° and γ = 120.0° | *P1* | 2.409Å | 2.303Å |

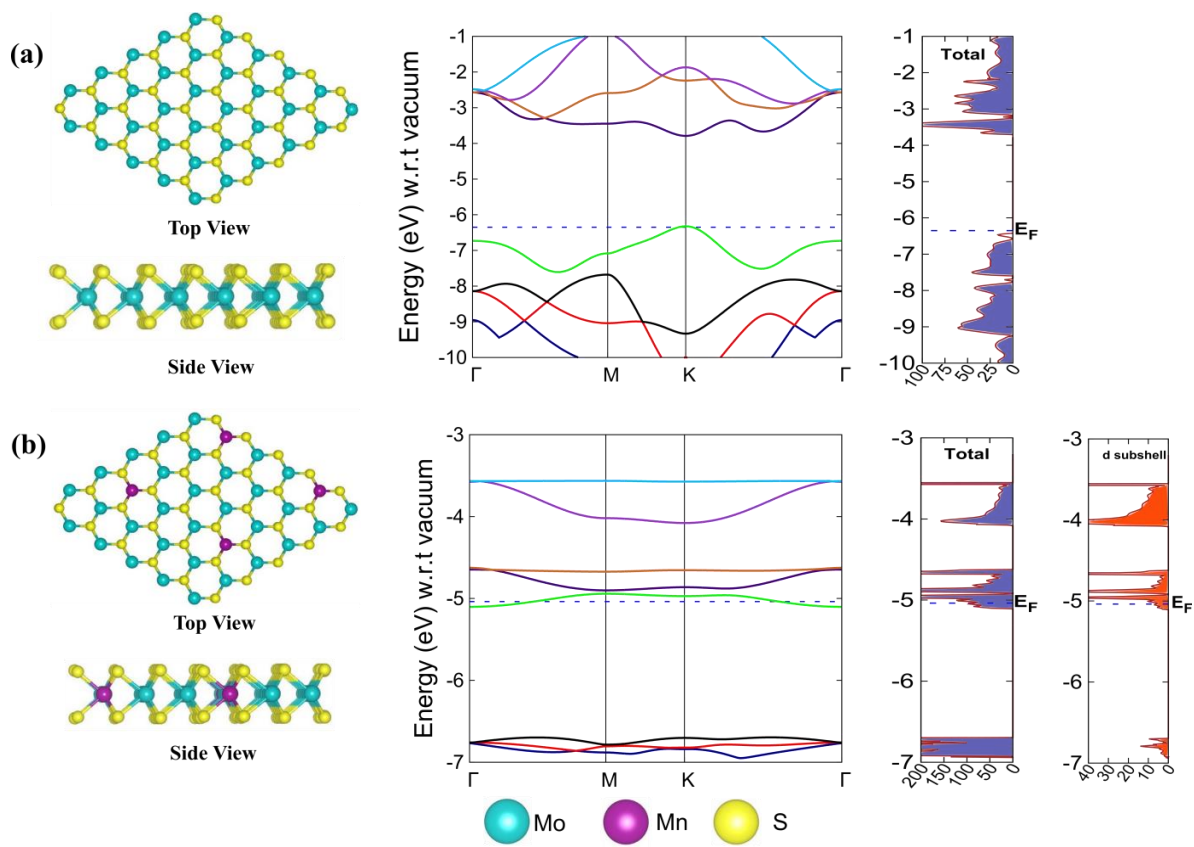

**Figure 2: (a)** Top view and side view of the pristine 2D monolayer MoS$_2$ with band structure and total DOS; **(b)** Top view and side view of the 2D monolayer Mn-MoS$_2$ with band structure and total DOS and the contributing component of the *d*-subshell DOS of the Mn atom in the total DOS.

Now it is confirmed from the electronic structure and properties calculations that the 2D Mn-MoS$_2$ material has shown the conducting properties (i.e. conducing in nature), so we proceed with our study in the direction of theoretical and computational development for optimum electrocatalyst. Hydrogen production from water electrolysis is conceived as a match for the growing need of alternating green energy source. In this regard, we carefully audit the fundamentals of HER and control parameters of the kinetics of reaction. We have outlined molecular simulation approaches that will help us to tackle the challenges that lie at hand in design of cheap and practical catalyst.

## 3.2 HER recapitulation

Now, we turned our attention to investigate the detailed HER mechanism by predicting energetics for the various reaction steps relevant for HER in the case of 2D monolayer Mn-MoS$_2$ material. Using the molecular cluster model system of the 2D Mn-MoS$_2$ material, we can now add or subtract electrons (e$^-$) and protons (H$^+$) independently in discrete H$_2$ evolution reaction steps. First, we calculate the free energies of the most likely intermediates to serve as a basis for describing the thermodynamics of HER. Then we examine the barriers of the various reaction steps to locate the rate limiting step during the reactions. In general, HER is two-way reaction mechanisms and the most generally accepted reaction mechanism is given as follows: HER can occur via the Volmer-Heyrovsky process or the Volmer-Tafel process as depicted in Figure 3. The process begins with water adsorption and dissociation where at first, the H$_2$O reacts with electron (e$^-$) to produce H$^+$ and OH$^-$ which takes place at the active site of the electrocatalyst (more specifically the active surface of the electrocatalyst). The further mechanism can occur either through Heyrovsky reaction step or the Tafel step. In the Heyrovsky reaction, the adsorbed hydride ion reacts with the adjacent water molecule or more specifically the with the solvated proton of the adjacent water to produce H$_2$. In the Tafel reaction step, where two adsorbed hydrogens are adjacent to each other, recombine to form H$_2$ during the reaction.

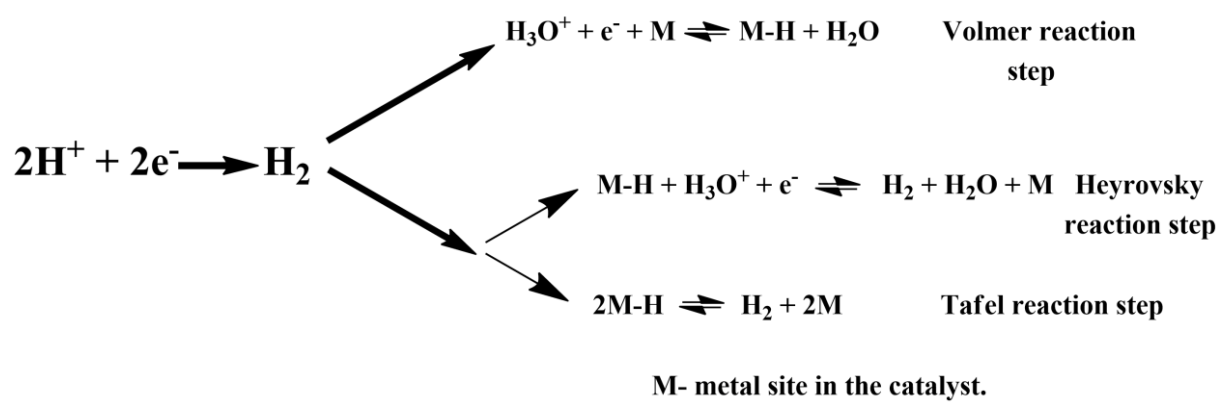

**Figure 3:** Possible reaction pathways of HER in acidic electrolyte.

Generally, it has been found that the Volmer-Heyrovsky reaction mechanism is more likely to be predominant when transition metal-based catalysts are used because of their good adsorption free energy.[71] An important fact arises in this mechanism is the activity descriptor of the mentioned mechanism. So, the analysis of the proposed catalyst for its involvement in the mechanics and kinetics of the reaction (as mentioned in Figure 3) was put into the effects. To study the electrocatalytic HER, we computationally developed a cluster model system for the 2D Mn-MoS$_2$ and performed non-periodic M06-L DFT theory.. The two processes have been carefully studied and discussed further.

## 3.3 Volmer-Heyrovsky Mechanism

The Volmer-Heyrovsky reaction pathway in the vicinity of the active site of 2D Mn-MoS$_2$ has been schematically presented in Figure 4. This process is a multistep electrode reaction which has been described, and the reaction steps, intermediates and transition states occurred during the HER process have been reported here.

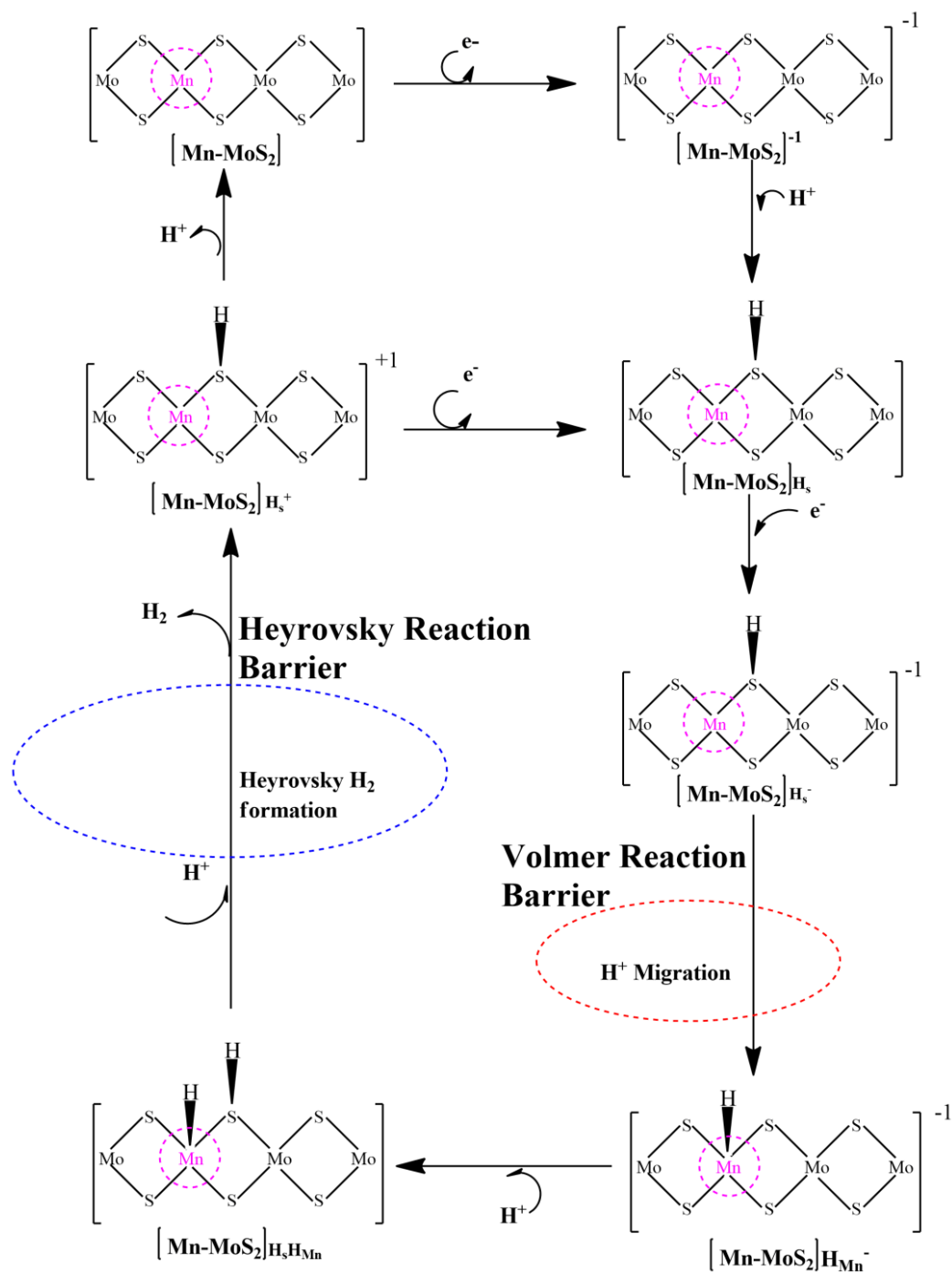

**Figure 4: Proposed reaction pathway scheme for HER using 2D monolayer Mn-MoS$_2$ electrocatalyst.**

In order to evolve an $H_2$ molecule, protons and electrons must be added to the 2D Mn-$MoS_2$ molecular cluster $Mn_1Mo_9S_{21}$. Here, it is useful to examine first the most stable structures with each number of extra electrons and each number of extra protons to understand the free energy differences between intermediate states, and ultimately find the lowest-barrier pathway. The detailed description of the HER process involved in the subject reaction is required to explain the electrochemistry. The first basic process is the dissociation of water.. To initiate the HER, one electron is absorbed on surface of the surface of the Mn-$MoS_2$ as depicted in Figure 4. This step takes place at standard hydrogen electrode (SHE) conditions which is the footing of thermodynamical potentials for oxidation and reduction processes. Thereafter $H^+$ from the medium is adsorbed on the sulfur site which is for the moment the most energetically conducive site, forming [Mn-$MoS_2$]$H_s$ solvated cluster. In the follow up step, [Mn-$MoS_2$]$H_s^{-1}$ complex is formed due to the addition of another electron from the solvent. Hence, we are reporting this HER as a two-electron transfer reaction. In the next step, the hydride ion from the sulfur site migrates to the next neighboring responsive Mn site. The migration of $H^{\bullet}$ from S to Mn is the Volmer reaction step. This transition structure i.e., the Volmer transition state (TS) is corroborated by the detection of imaginary vibrational frequency at the site of transition of hydride ion from S to Mn. The formation of TS is accompanied by positive free energy change of 11.8 kcal mol$^{-1}$. $H^+$ from medium again attacks the S site and gets attached to surface S to form [Mn-$MoS_2$]$H_sH_{Mn}$ cluster. As discussed earlier, from here either the Tafel or the Heyrovsky process can take place. The Heyrovsky part is further shown in Figure 4. Computationally we explicitly added 4$H_2O$-$H^+$ cluster in vicinity of the active site of the catalyst and observed after the simulation that $H^-$ from Mn and the $H^+$ from the water cluster evolve as $H_2$. This process is called Heyrovsky process, often mentioned as the Heyrovsky transition state (TS2) in reaction mechanism during HER. The Mn-S bond distance during TS2 was recorded to be 2.311 Å which is higher than the strain free bond distance of 2.303 Å as mentioned in Table 1. All the optimized reaction intermediates with TSs of our proposed reaction scheme have been illustrated in Figure 5.

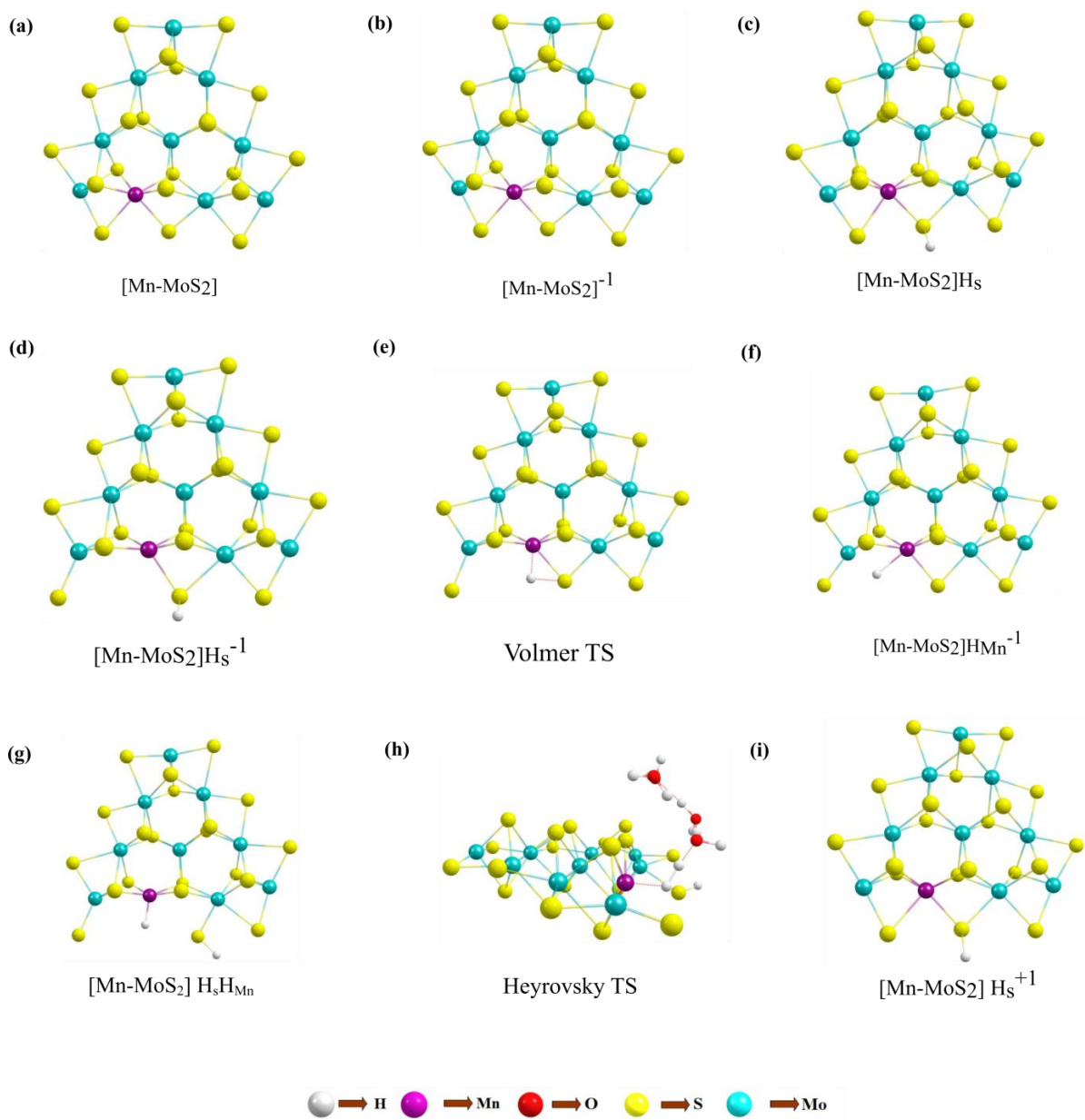

**Figure 5: The equilibrium geometries of the molecular model systems of the 2D Mn-MoS$_2$ materials have been considered in the subject reaction.**

We keenly followed the different reaction pathway schemes as shown in Figure 4 but kept our focus on the two important saddle points i.e., Volmer reaction where an H• atom migrates from S to the transition metal site and the other being Heyrovsky reaction step where H$^+$ from the adjacent water cluster with the hydronium and the H$^-$ from the Mn site combine to form H$_2$. It has been observed that these two reaction steps are the rate determining steps of the HER. The free energy, enthalpy and total energy change at each reaction step is listed in

Table 2. From our gas phase calculations, we observed the Volmer activation barrier is about 7.23 kcal mol$^{-1}$ and the Heyrovsky reaction barrier is about 10.60 kcal mol$^{-1}$ in the 2D Mn-MoS$_2$ materials.

**Table 2:** Energy changes during reaction mechanism in gas phase.

| Reaction Intermediates | ΔE (kcal mol$^{-1}$) | ΔH (kcal mol$^{-1}$) | ΔG (kcal mol$^{-1}$) |
|---|---|---|---|
| [Mn-MoS$_2$] ⟶ [Mn-MoS$_2$]$^{-1}$ | -30.05 | -30.09 | -30.12 |
| [Mn-MoS$_2$]$^{-1}$ ⟶ [Mn-MoS$_2$]H$_s$ | -26.98 | -27.21 | -27.25 |
| [Mn-MoS$_2$]H$_s$ ⟶ [Mn-MoS$_2$]H$_s^{-1}$ | -29.59 | -30.04 | -31.73 |
| [Mn-MoS$_2$]H$_s^{-1}$ ⟶ Volmer TS | 8.17 | 8.02 | 7.23 |
| Volmer TS ⟶ [Mn-MoS$_2$]H$_{Mn}^{-1}$ | 12.76 | -12.94 | -13.04 |
| [Mn-MoS$_2$]H$_{Mn}^{-1}$ ⟶ [Mn-MoS$_2$]H$_s$H$_{Mn}$ | -58.26 | -58.47 | -58.65 |
| [Mn-MoS$_2$]H$_s$H$_{Mn}$ ⟶ Heyrovsky TS | 10.73 | 10.64 | 10.60 |
| Heyrovsky TS ⟶ [Mn-MoS$_2$] H$_s^{+1}$ | -60.78 | -60.79 | -61.13 |

**Table 3 :** Reaction Energy Barriers in HER when Mn-MoS$_2$ catalyst is used.

| Activation Barrier | ΔG (kcal mol$^{-1}$) in gas phase | ΔE (kcal mol$^{-1}$) in solvent phase |
|---|---|---|
| **Volmer reaction barrier** | 7.23 | 11.84 |
| **Heyrovsky reaction barrier** | 10.60 | 13.46 |

From our DFT-D calculations, we have predicted that the Volmer reaction energy barrier in the solvent phase is about as 11.84 kcal mol$^{-1}$. Recently, Yu et. al.[4] reported that this energy barrier for the pristine MoS$_2$, WS$_2$ and hybrid W$_{0.4}$Mo$_{0.6}$S$_2$ alloy is 17.7 kcal.mol$^{-1}$, 18.1

kcal.mol$^{-1}$ and 11.9 kcal.mol$^{-1}$, respectively, in the solvent phase. Moreover, the Heyrovsky barrier reported for the pristine MoS$_2$, WS$_2$ and hybrid W$_{0.4}$Mo$_{0.6}$S$_2$ alloy were calculated as 23.8 kcal mol$^{-1}$, 21.3 kcal mol$^{-1}$ and 13.3 kcal mol$^{-1}$, respectively.. In the case of the 2D monolayer Mn-MoS$_2$ material, the calculated Heyrovsky energy barrier is about 13.45 kcal mol$^{-1}$ in the solvent phase. Both the Volmer and Heyrovsky reaction energy barriers for the different materials are mentioned in Table 4. We reported from our DFT calculations that our proposed 2D monolayer Mn-MoS$_2$ catalyst shows much lower reaction barriers during the HER, thus, the 2D monolayer Mn-MoS$_2$ can be considered as a highly efficient HER electrocatalyst.

**Table 4:** Comparison of the Volmer and Heyrovsky reaction energy barriers during the HER in the solvent phase using different catalyst is reported here.

| Material | Volmer Reaction Barrier (kcal mol$^{-1}$) | Heyrovsky Reaction Barrier (kcal mol$^{-1}$) | References |
|---|---|---|---|
| MoS$_2$ | 17.7 | 23.8 | 4 |
| WS$_2$ | 18.1 | 21.3 | 4 |
| W$_{0.4}$Mo$_{0.6}$S$_2$ | 11.9 | 13.3 | 4 |
| Mn-MoS$_2$ | 11.8 | 13.5 | This work |

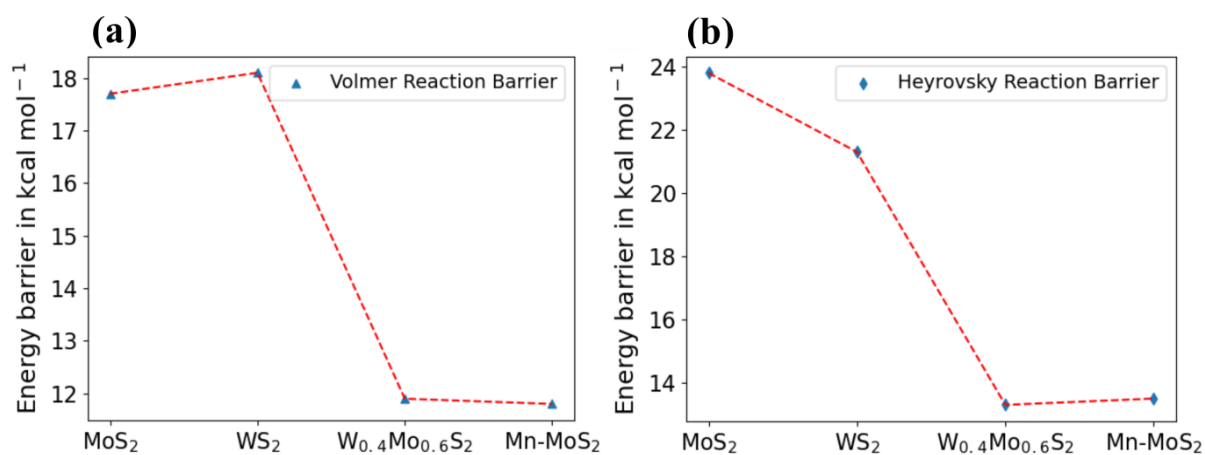

**Figure 6: Graphical illustration of the (a) Volmer and (b) Heyrovsky reaction barriers in the solvent phase of the 2D monolayer MoS2, WS2, W$_{0.4}$Mo$_{0.6}$S$_2$ and Mn-MoS$_2$ materials are shown here.**

From the transition state theory (TST)[72] or the activated complex theory by including the DFT calculations, we have determined the turnover frequency (TOF) for H$_2$ evolution per edge of the doped Mn atom in the 2D Mn-MoS$_2$ catalyst. For the theoretical determination, we used the formula : rate = (K$_B$T/h) × exp(-ΔG/RT);[73] where K$_B$ is the Boltzmann constant, T is temperature, h is the Planck's constant, R is the universal gas constant and ΔG corresponds to the energy barrier. The TOF obtained for the 2D monolayer Mn-MoS$_2$ material from H$_2$ formation reaction energy barrier in the Heyrovsky mechanism (in the solvent phase) is about $8.48 \times 10^2$ sec$^{-1}$. The high value of the TOF is suitable for better performance of the H$_2$ evolution during the reaction.

**Table 5 :** Heyrovsky reaction barrier and TOF for the pristine MoS$_2$, WS$_2$, W$_{0.4}$Mo$_{0.6}$S$_2$ alloy and the Mn-MoS$_2$ materials have been computed in the solvent phase at the DFT level of theory..

| Material | Barrier in gas phase (kcal mol$^{-1}$) | Barrier in solvent phase (kcal mol$^{-1}$) | Turnover frequency in solvent phase (sec$^{-1}$) | Reference |
|---|---|---|---|---|
| **MoS$_2$** | 16.0 | 23.8 | $2.1 \times 10^{-5}$ | 4 |
| **WS$_2$** | 14.5 | 21.3 | $1.5 \times 10^{-3}$ | 4 |
| **W$_{0.4}$M$_{0.6}$S$_2$** | 11.5 | 13.3 | $1.1 \times 10^3$ | 4 |
| **Mn-MoS$_2$** | 10.6 | 13.5 | $8.4 \times 10^2$ | This work |

The comparison of the activation energies and TOF of different materials as depicted in Table 1 provide insight of their catalytic activity. We can observe that 2D monolayer Mn-MoS$_2$ materials shows comparable results to for the hybrid W$_{0.4}$Mo$_{0.6}$S$_2$ material. Therefore, it can be mentioned here that the 2D monolayer Mn-MoS$_2$ material can prove to be a better and practical alternative for superb catalytic performance for HER. Another electrochemical parameter i.e., Tafel slope can also be calculated theoretically by taking into consideration the number of electron transfer during HER mechanism. As stated earlier that the proposed reaction is a two-electron transfer mechanism, it has been found that the computed Tafel slope turned out to be 29.55 mV dec$^{-1}$ for the 2D Mn-MoS$_2$.

Our present computations are in strong favor of low energy barriers in both the Volmer and Heyrovsky steps during the HER on the surfaces of the 2D monolayer Mn-MoS$_2$ resulting in a promising candidate for HER as an electrocatalyst. To further support our development, we implemented the Natural Bond Orbital (NBO), highest occupied molecular orbital (HOMO) and lowest unoccupied molecular orbital (LUMO) calculations at the equilibrium structures using the same DFT method. These calculations were performed in order to show appropriate perspective of H$_2$ formation at the active site from the electronic charge and molecular orbital overlapping point of view. Precise Lewis structures i.e., structures which have maximum electronic charge in the Lewis orbitals, can be found out by calculating NBOs. This study conveys interaction density or the overlap density from the wavefunctions. The solution to the multi electron atomic system requires an approximation called the linear combination of atomic orbitals (LCAO approximation). The qualitative picture of molecular orbital is analyzed by expanding the molecular orbital into any complete basis set of all atomic orbitals of nuclei. So, the multi-electron wavefunction in a molecule at a specific configuration of the nuclei can be given by expanding the orbital approximation to molecules. The wave function obtained from the NBO calculations is a linear combination of the atomic orbitals of the Mn, S, Mo and H for the Volmer TS and Mn, S, Mo, H and O for the Heyrovsky TS. The HOMO-LUMO was obtained from the optimized transition (both Volmer and Heyrovsky) structures and has been shown in Figure 7a-d. The red color represents in phase bonding of the orbitals and the blue color shows out of phase bonding. The boundary value outlining the isosurface shown in Figure 7 was set at 0.009. The interval of values in which the isosurface is colored (from blue to red) was set from -0.1 to 0.1.

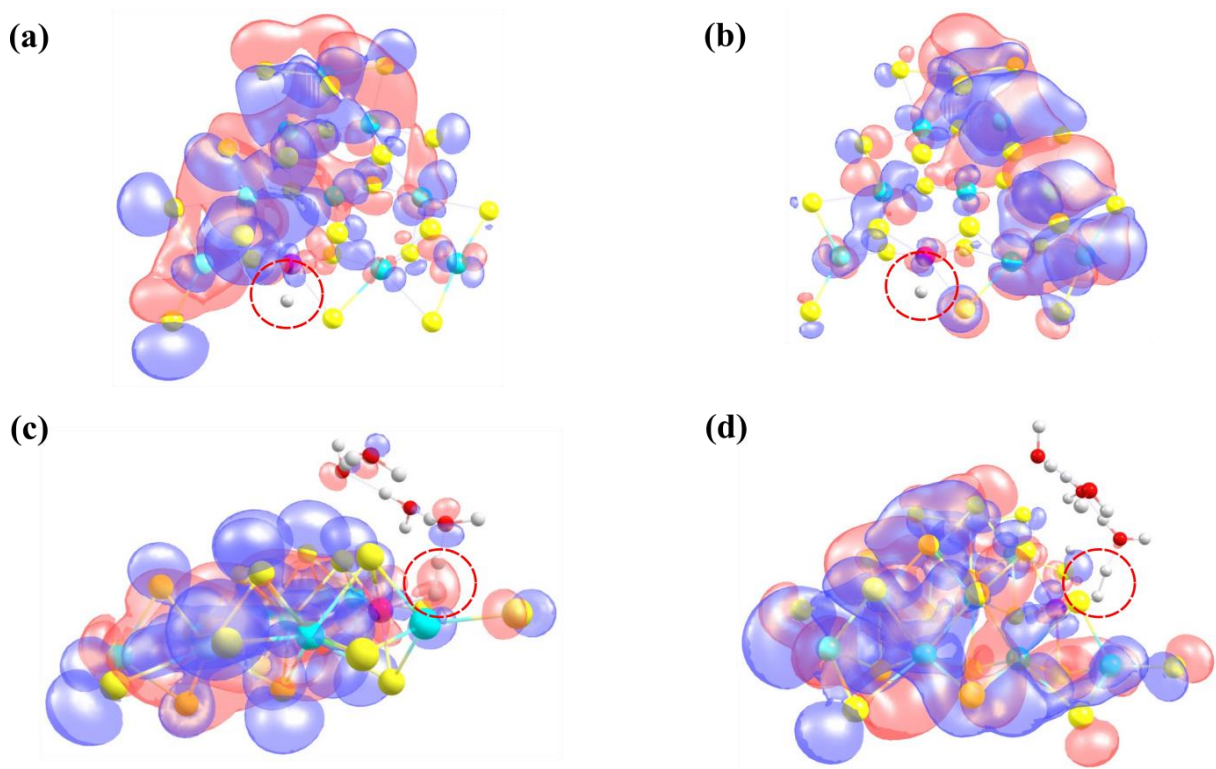

**Figure 7:** The equilibrium structures of (a) the HOMO of the Volmer TS; (b) the LUMO of the Volmer TS; (c) the HOMO of the Heyrovsky TS; and (d) the LUMO of the Heyrovsky TS have been displayed here. The molecular orbitals involved in the subject reaction and the position of hydrogen have been highlighted by red dotted circle.

The insight can be drawn on the role of electronic structure in the HER mechanism from the HOMO-LUMO calculations of the Heyrovsky transition state i.e. TS2, and the $H_2$ formation in a steady state due to better overlap of the *d*-orbital of the Mn atom and the *s*-orbital of $H_2$. Therefore, one conclusion can be drawn such that in the rate limiting step of HER i.e., the Heyrovsky step, the stabilization of the atomic orbitals is also one of the key features for reducing this reaction barrier. The electron cloud around the H atoms in Heyrovsky TS2 is highlighted by red dotted circle (Figure 7 (c)). This step is backed up by the overlap of the atomic orbitals of the H and Mn atoms along with $H_3O^+$ ion when the $H_2$ is evolved. This is also one of the reasons why the 2D monolayer Mn-$MoS_2$ shows excellent activity for HER. The energy difference between the HOMO and LUMO, also known as HOMO-LUMO gap is used to predict the stability transition metal based complex[74] as it is the lowest energy electronic excitation that is possible in the molecule.

## 3.4 Volmer-Tafel mechanism

The proposed Volmer-Tafel reaction scheme is illustrated in Figure 8.

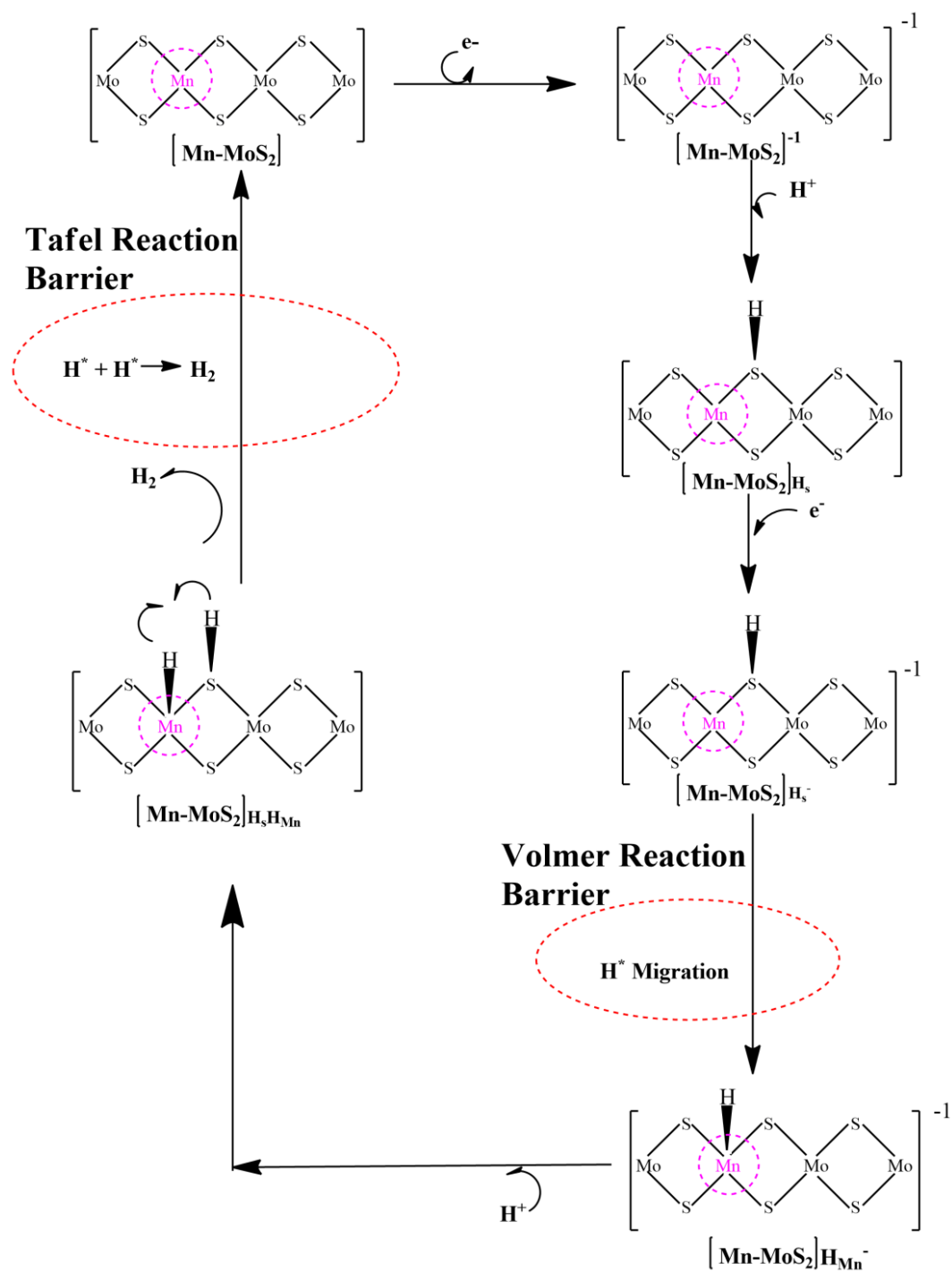

**Figure 8:** The detailed reaction scheme of the Volmer – Tafel mechanism is depicted here.

The reaction steps are similar till the formation of the [Mn-MoS$_2$]H$_s$H$_{Mn}$ complex during the subject reaction. In this Volmer-Tafel mechanism, the process takes place as follows; two adsorbed hydrogen atoms (here H$_s$ and H$_{Mn}$) on the surface of the catalyst combine to evolve as H$_2$. The equilibrium structure of the Volmer-Tafel transition state (TS3) is shown in Figure

9a.

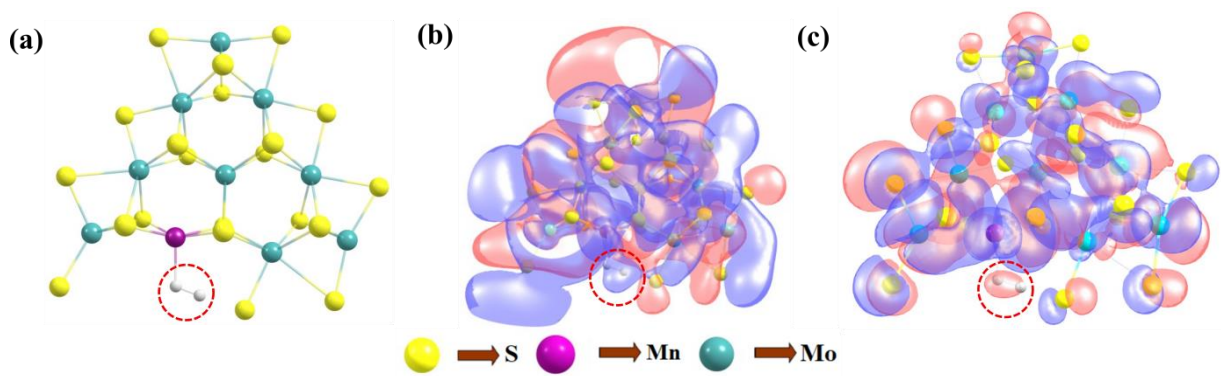

**Figure 9: (a) The equilibrium Volmer-Tafel transition state TS3; (b) HOMO and (c) LUMO of the Volmer-Tafel TS have been depicted here.**

In this reaction the adsorbed hydrogen at sulfur site and the adsorbed hydrogen at the transition metal site (here Mn-site) combine to form $H_2$ ($2H^* \rightarrow H_2$; where $H^*$ represents hydrogen adsorbed on the active site of the catalyst). The Tafel barrier reaction was recorded to be 90.13 kcal mol$^{-1}$ in the gas phase and 93.66 kcal mol$^{-1}$ in the solvent phase which indicates a high energy barrier for the Volmer-Tafel reaction mechanism. The HOMO-LUMO calculations were also performed to better visualize the Tafel reaction mechanism as depicted in Figure 9(b) and 9(c). The electron cloud represents both positive and negative part of the wavefunction by red and blue color. The electron cloud around hydrogen is highlighted by red dotted circle. In Figure 9b which represents the HOMO of Tafel TS, the orbitals around $H_2$ formed during TS is blue. This means that the orbital mixing is out of phase. The red cloud around $H_2$ in LUMO of Tafel TS suggests in-phase interaction of electronic wavefunctions. The phase or orbital is a direct consequence of the wave like property of electrons and generally the in-phase mixing suggests lower energy state and the out of phase mixing indicates anti-bonding orbitals or higher energy state. The corresponding TOF in gas phase and solvent phase was calculated to be $6.19 \times 10^{-45}$ sec$^{-1}$ and $4.59 \times 10^{-50}$ sec$^{-1}$ respectively. This TOF value is very low, and hence the process is least likely to take place. It should be noted here that the Volmer-Tafel reaction barrier is much higher than the calculated Heyrovsky barrier as depicted in Table 6. Hence, the Volmer-Heyrovsky reaction will be more assertive than the Volmer-Tafel reaction when the 2D monolayer Mn-MoS$_2$ material can be used as an effective electrocatalyst for HER. Heteroatom doping in the pristine 2D monolayer MoS$_2$ has led to a significant change in electronic properties of the pristine TMD materials. As shown in our present computed results, the 2D monolayer Mn-MoS$_2$ shows excellent electrocatalytic performance. The results of the descriptor-based method aided by the DFT computations have been thoroughly discussed above. This indicates that 2D monolayer Mn-MoS$_2$ driven catalysis is a viable and efficient hydrogen production method.

**Table 6: All reaction barriers in HER mechanism using 2D Mn-MoS$_2$**

| Activation Barrier | ΔG (kcal mol$^{-1}$) in gas phase | ΔG (kcal mol$^{-1}$) in solvent phase |
|---|---|---|
| **Volmer reaction barrier** | 7.23 | 11.84 |
| **Heyrovsky reaction barrier** | 10.60 | 13.46 |
| **Tafel reaction barrier** | 90.13 | 93.66 |

# 4. Conclusion

In summary, a theoretical development has been proposed for an effective HER electrocatalyst where 2D monolayer Mn doped MoS$_2$ actively demonstrates promising results. In this comprehensive study, we have encapsulated the relationship between the structure and morphology of the material that characterizes its catalytic activity. Lowering of the activation barrier is one of the key features of the catalyst and the electronic overlap between the *s*-orbital of hydrogen and the *d*-orbitals of the transition metal in TMD has favored to H$_2$ formation. The extensive study of the 2D monolayer Mn-MoS$_2$ can play a key role in electrocatalytic HER step for H$_2$ production as a low cost and non-noble metal based electrocatalyst. The nature of the reaction is complex which further complicates the experimental investigation whose conclusion is collateral and dependent mostly upon the sample. Hence, we used the density functional theory method in the present study to determine the properties of 2D Mn-MoS$_2$, and the influence on thermodynamics of the reaction by scrutinizing at length the computational model at the edge of the structure. The quantum mechanical based DFT computational approach has helped to describe the HER mechanism at the atomistic level. The hybrid periodic DFT-D method showed that the 2D layer of Mn-MoS$_2$ has zero band gap, and the total DOS calculations showed that it became electron rich due to the addition of Mn in the MoS$_2$. The examination of the performance of the 2D monolayer Mn-MoS$_2$ material for catalytic activity has been done through Mn$_1$Mo$_9$S$_{21}$ molecular cluster model. The detailed reaction mechanism along with the transition states has been calculated by M06-L DFT method. The reaction followed two electron transfer kinetics with highly favorable Volmer-Heyrovsky mechanism. The Volmer and Heyrovsky barrier were 11.5 kcal mol$^{-1}$ and 13.5 kcal mol$^{-1}$, respectively, in the solvent phase computed by the same DFT method. The rate determining step has been stabilized because of the better overlap of the atomic orbitals i.e., the *d*-orbital of Mn and the *s*-orbital of H$_2$ molecule. Although HER can also follow Volmer-Tafel mechanism but the Tafel transition barrier was found to be 87.18 kcal mol$^{-1}$ so the Volmer-Tafel reaction is less likely to occur on the surfaces of the 2D Mn-MoS$_2$ material. High TOF, lower activation barriers and the theoretically determined Tafel slope (29.55 mV dec$^{-1}$), all attributed to the Mn-MoS$_2$ being a promising and efficient electrocatalyst for HER. In eventuality the catalytic activity of 2D TMDs can be coordinated by activating the basal plane. So immense potential can be found in layered heterostructures for electrocatalytic activity in HER and this type of study will further trigger the research in this direction.

## Author Contributions:

Dr Pakhira developed the complete idea of this current research work, and he computationally studied the electronic structures and properties of the 2D TMDs MoS$_2$ and Mn-MoS$_2$. Dr Pakhira explored the whole reaction pathways; transitions states and reactions barriers and he explained the HER mechanism by the DFT calculations. Quantum calculations and theoretical models were designed and performed by Dr Pakhira and Mr. Joy Ekka, and Mr. Shrish Nath Upadhyay and Mr. Verma Bunty Sardar helped to Mr. Joy Ekka to perform the DFT calculations in the high-performance computing system. Dr Pakhira and Mr. Joy Ekka wrote the whole manuscript and prepared all the tables and figures in the manuscript. Mr. Shrish Nath Upadhyay and Mr. Verma Bunty Sardar helped Dr Pakhira to organize the manuscript.

# AUTHOR INFORMATION

**Corresponding Author**
**Dr. Srimanta Pakhira** − *Discipline of Physics, Indian Institute of Technology Indore (IITI), Indore, MP 453552, India;*
*Discipline of Metallurgy Engineering and Materials Science, Indian Institute of Technology Indore (IITI), Indore, MP 453552, India;*
*Centre of Advanced Electronics (CAE), Indian Institute of Technology Indore, Indore, MP 453552, India;*
ORCID: orcid.org/0000-0002-2488-300X;
Email: spakhira@iiti.ac.in or spakhirafsu@gmail.com

**Authors**

**Joy Ekka** − *Discipline of Physics, Indian Institute of Technology Indore (IITI), Indore, MP 453552, India*

**Shrish Nath Upadhyay** − *Discipline of Metallurgy Engineering and Materials Science (MEMS), Indian Institute of Technology Indore (IITI), Indore, MP 453552, India*;
ORCID: orcid.org/0000-0003-0029-4160.

**Verma Bunty Sardar** − *Discipline of Physics, Indian Institute of Technology Indore (IITI), Indore, MP 453552, India*

## Acknowledgment:

This work was financially supported by the Science and Engineering Research Board-Department of Science and Technology (SERB-DST), Government of India under the Grant No. ECR/2018/000255. Dr. Srimanta Pakhira thanks the Science and Engineering Research Board, Department of Science and Technology (SERB-DST), Govt. of India for providing his highly prestigious Ramanujan Faculty Fellowship under the scheme no. SB/S2/RJN-067/2017, and for his Early Career Research Award (ECRA) under the grant No. ECR/2018/000255. Mr. Upadhyay thanks Indian Institute of Technology Indore, MHRD, Govt. of India for providing the doctoral fellowship. The authors would like to acknowledge Indian Institute of Technology Indore (IITI) for providing the basic infrastructure to conduct this research work.

# References

1 S. E. Hosseini and M. A. Wahid, *Renew. Sustain. Energy Rev.*, 2016, **57**, 850–866.

2 R. De Levie, 1999, **476**, 92–93.

3 N. Cheng, S. Stambula, D. Wang, M. N. Banis, J. Liu, A. Riese, B. Xiao, R. Li, T. K. Sham, L. M. Liu, G. A. Botton and X. Sun, *Nat. Commun.*, 2016, **7**, 1–9.

4 Y. Lei, S. Pakhira, K. Fujisawa, X. Wang, O. O. Iyiola, N. Perea López, A. Laura Elías, L. Pulickal Rajukumar, C. Zhou, B. Kabius, N. Alem, M. Endo, R. Lv, J. L. Mendoza-Cortes and M. Terrones, *ACS Nano*, 2017, **11**, 5103–5112.

5 K. Liang, S. Pakhira, Z. Yang, A. Nijamudheen, L. Ju, M. Wang, G. E. Sterbinsky, Y. Du, Z. Feng, J. L. Mendoza-cortes and Y. Yang, *ACS Catal.*, 2018, **1**, 651–659.

6 S. Z. Qiao, Y. Zheng, Y. Jiao, M. Jaroniec and S. Z. Qiao, *Angew. Chemie Int. Ed.*, 2014, **54**, 52–65.

7 X. Zheng, J. Xu, K. Yan, H. Wang, Z. Wang and S. Yang, *Chem. Mater.*, 2014, **26**, 2344–2353.

8 P. Liu and J. A. Rodriguez, *J. Am. Chem. Soc.*, 2005, **127**, 14871–14878.

9 A. K. Singh, J. Prasad, U. P. Azad, A. K. Singh, R. Prakash, K. Singh, A. Srivastava, A. A. Alaferdov and S. A. Moshkalev, *RSC Adv.*, 2019, **9**, 22232–22239.

10 A. E. Russell, *Faraday Discuss.*, 2008, **140**, 9–10.

11 Y. P. Venkata Subbaiah, K. J. Saji and A. Tiwari, *Adv. Funct. Mater.*, 2016, **26**, 2046–2069.

12 Y. Li, H. Wang, L. Xie, Y. Liang, G. Hong and H. Dai, *J. Am. Chem. Soc.*, 2011, **133**, 7296–7299.

13 D. Kong, H. Wang, J. J. Cha, M. Pasta, K. J. Koski, J. Yao and Y. Cui, *Nano Lett.*, 2013, **13**, 1341–1347.

14 J. Yang and H. S. Shin, *J. Mater. Chem. A*, 2014, **2**, 5979–5985.

15  Y. Yan, B. Xia, Z. Xu and X. Wang, *ACS Catal.*, 2014, **4**, 1693–1705.

16  R. Ye, P. Del Angel-Vicente, Y. Liu, M. J. Arellano-Jimenez, Z. Peng, T. Wang, Y. Li, B. I. Yakobson, S. H. Wei, M. J. Yacaman and J. M. Tour, *Adv. Mater.*, 2016, **28**, 1427–1432.

17  A. Nipane, D. Karmakar, N. Kaushik, S. Karande and S. Lodha, *ACS Nano*, 2016, **10**, 2128–2137.

18  X. Shang, W. H. Hu, X. Li, B. Dong, Y. R. Liu, G. Q. Han, Y. M. Chai and C. G. Liu, *Electrochim. Acta*, 2017, **224**, 25–31.

19  Y. Yin, J. Han, Y. Zhang, X. Zhang, P. Xu, Q. Yuan, L. Samad, X. Wang, Y. Wang, Z. Zhang, P. Zhang, X. Cao, B. Song and S. Jin, *J. Am. Chem. Soc.*, 2016, **138**, 7965–7972.

20  T. F. Jaramillo, K. P. Jørgensen, J. Bonde, J. H. Nielsen, S. Horch and I. Chorkendorff, *Science (80-. ).*, 2007, **317**, 100–102.

21  A. D. Becke and A. D. Becke, *J. Chem. Phys.*, 1993, **98**, 5648–5652.

22  S. Pakhira, T. Debnath, K. Sen and A. K. Das, *J. Chem. Sci.*, 2016, **128**, 621–631.

23  S. Pakhira and J. L. Mendoza-Cortes, *Phys. Chem. Chem. Phys.*, 2019, **21**, 8785–8796.

24  S. Pakhira and J. L. Mendoza-Cortes, *J. Phys. Chem. C*, 2018, **122**, 4768–4782.

25  W. Niu, S. Pakhira, K. Marcus, Z. Li and J. L. Mendoza-cortes, *Adv. Energy Mater.*, 2018, **8**, 1800480.

26  S. Pakhira, *RSC Adv.*, 2019, **9**, 38137–38147.

27  A. J. Freeman, *J. Comput. Appl. Math.*, 2002, **149**, 27–56.

28  M. J. Frisch, G. W. Trucks, H. B. Schlegel, G. E. Scuseria, M. A. Robb, J. R. Cheeseman, G. Scalmani, V. Barone, G. A. Petersson, H. Nakatsuji, X. Li, M. Caricato, A. V. Marenich, J. Bloino, B. G. Janesko, R. Gomperts, B. Mennucci and J. B. Hratch, 2016, Gaussian 16 (Revision C.01) Gaussian, Inc., Wallin.

29  A. D. Becke, *J. Chern. Phys.*, 2005, **98**, 5648–5652.

30  E. Caldeweyher, C. Bannwarth, S. Grimme, E. Caldeweyher, C. Bannwarth and S.

Grimme, *J. Chem. Phys.*, 2017, **147**, 034112.

31  A. Hansen, C. Bauer, S. Ehrlich, A. Najibi and S. Grimme, *Phys. Chem. Chem. Phys.*, 2017, **19**, 32184–32215.

32  S. Grimme, J. Antony, S. Ehrlich, H. Krieg, S. Grimme, J. Antony, S. Ehrlich and H. Krieg, *J. Chem. Phys.*, 2010, **132**, 154104.

33  S. Pakhira, M. Takayanagi and M. Nagaoka, *J. Phys. Chem. C*, 2015, **119**, 28789–28799.

34  S. Pakhira, K. P. Lucht and J. L. Mendoza-Cortes, *J. Phys. Chem. C*, 2017, **121**, 21160–21170.

35  S. Pakhira, K. Sen, C. Sahu and A. K. Das, *J. Chem. Phys.*, 2013, **138**, 164319.

36  N. Sinha and S. Pakhira, *ACS Appl. Electron. Mater.*, 2021, **3**, 720−732.

37  R. Puttaswamy, R. Nagaraj, P. Kulkarni, H. K. Beere, S. N. Upadhyay, R. G. Balakrishna, N. Sanna Kotrappanavar, S. Pakhira and D. Ghosh, *ACS Sustain. Chem. Eng.*, 2021, **9**, 3985–3995.

38  J. Hui, S. Pakhira, R. Bhargava, Z. J. Barton, X. Zhou, A. J. Chinderle, J. L. Mendoza-Cortes and J. Rodríguez-López, *ACS Nano*, 2018, **12**, 2980–2990.

39  R. Dovesi, A. Erba, R. Orlando, C. M. Zicovich-Wilson, B. Civalleri, L. Maschio, M. Rérat, S. Casassa, J. Baima, S. Salustro and B. Kirtman, *Wiley Interdisc. Rev. Comput. Mol. Sci.*, 2018, **8**, 1–36.

40  S. Pakhira and J. L. Mendoza-Cortes, *arXiv*, 2019, **124**, 6454−6460.

41  K. Sen, S. Pakhira, C. Sahu and A. K. Das, *Mol. Phys.*, 2014, **112**, 182–188.

42  S. Pakhira, T. Debnath, K. Sen and A. K. Das, *J. Chem. Sci.*, 2016, **128**, 621–631.

43  S. Pakhira, C. Sahu, K. Sen and A. K. Das, *Struct. Chem.*, 2013, **24**, 549–558.

44  S. Pakhira and A. K. Dasa, *Eur. Phys. J. D*, 2012, **66**, 144–154.

45  S. Pakhira, B. Mondal and A. K. Das, *Chem. Phys. Lett.*, 2011, **505**, 81–86.

46  D. Vilela Oliveira, J. Laun, M. F. Peintinger and T. Bredow, *J. Comput. Chem.*, 2019, **40**, 2364–2376.

47    M. F. Peintinger, D. V. Oliveira and T. Bredow, *J. Comput. Chem.*, 2013, **34**, 451–459.

48    C. Fisica and V. Giuria, 1997, **7**, 959–967.

49    A. Montoya, T. N. Truong and A. F. Sarofim, *J. Phys. Chem. A*, 2000, **104**, 6108–6110.

50    S. Pakhira and J. L. Mendoza-Cortes, *J. Phys. Chem. C*, 2020, **124**, 6454–6460.

51    S. Pakhira, B. S. Lengeling, O. Olatunji-Ojo, M. Caffarel, M. Frenklach and W. A. Lester, *J. Phys. Chem. A*, 2015, **119**, 4214–4223.

52    J. Baker, A. Scheiner and J. Andzelm, *Chem. Phys. Lett.*, 1993, **216**, 380–388.

53    S. Pakhira, K. P. Lucht and J. L. Mendoza-cortes, *J. Chem. Phys.*, 2018, **148**, 064707.

54    K. Hu, M. Wu, S. Hinokuma, T. Ohto, M. Wakisaka, J. I. Fujita and Y. Ito, *J. Mater. Chem. A*, 2019, **7**, 2156–2164.

55    K. Momma and F. Izumi, *J. Appl. Crystallogr.*, 2011, 44, 1272–1276.

56    Y. Zhao and D. G. Truhlar, *J. Chem. Phys.*, 2006, **125**, 194101.

57    Y. Zhao and D. G. Truhlar, *Theor. Chem. Acc.*, 2008, **120**, 215–241.

58    A. J. Garza, S. Pakhira, A. T. Bell, J. L. Mendoza-Cortes and M. Head-Gordon, *Phys. Chem. Chem. Phys.*, 2018, **20**, 24058–24064.

59    S. Grimme, T. O. Chemie and O. I. D. U. Münster, 2006, **27**, 1787–1799.

60    R. Ditchfield, W. J. Hehre and J. A. Pople, *J. Chem. Phys.*, 1971, **54**, 720–723.

61    P. C. Hariharan and J. A. Pople, *Theor. Chim. Acta*, 1973, **28**, 213–222.

62    M. M. Francl, W. J. Pietro, W. J. Hehre, J. S. Binkley, M. S. Gordon, D. J. DeFrees and J. A. Pople, *J. Chem. Phys.*, 1982, **77**, 3654–3665.

63    A. You, M. A. Y. Be and I. In, *J. Chem. Phys.*

64    V. A. Rassolov, J. A. Pople, M. A. Ratner and T. L. Windus, *J. Chem. Phys.*, 1998, **109**, 1223–1229.

65    P. J. Hay and W. R. Wadt, *J. Chem. Phys.*, 1985, **82**, 270–283.

66   P. J. Hay and W. R. Wadt, *J. Chem. Phys.*, 1985, **82**, 299–310.

67   Chemcraft - graphical software for visualization of quantum chemistry computations, https://www.chemcraftprog.com.

68   F. Lipparini and B. Mennucci, *J. Chem. Phys.*, 2016, **144**, 160901.

69   X. L. Fan, Y. R. An and W. J. Guo, *Nanoscale Res. Lett.*, 2016, **11**, 154–163.

70   J. Ryou, Y. S. Kim, K. C. Santosh and K. Cho, *Sci. Rep.*, 2016, **6**, 1–8.

71   C. Li, H. Gao, W. W. Wan and T. Mueller, *Phys. Chem. Chem. Phys.*, 2019, **21**, 24489–24498.

72   P. C. Jordan and P. C. Jordan, *Chem. Kinet. Transp.*, 1979, 269–323.

73   Y. Huang, R. J. Nielsen, W. A. Goddard and M. P. Soriaga, *J. Am. Chem. Soc.*, 2015, **137**, 6692–6698.

74   R. Srinivasaraghavan, S. Thamaraikannan, S. Seshadri and T. Gnanasambandan, *Spectrochim. Acta - Part A Mol. Biomol. Spectrosc.*, 2015, **137**, 1194–1205.

**Graphical Abstract**

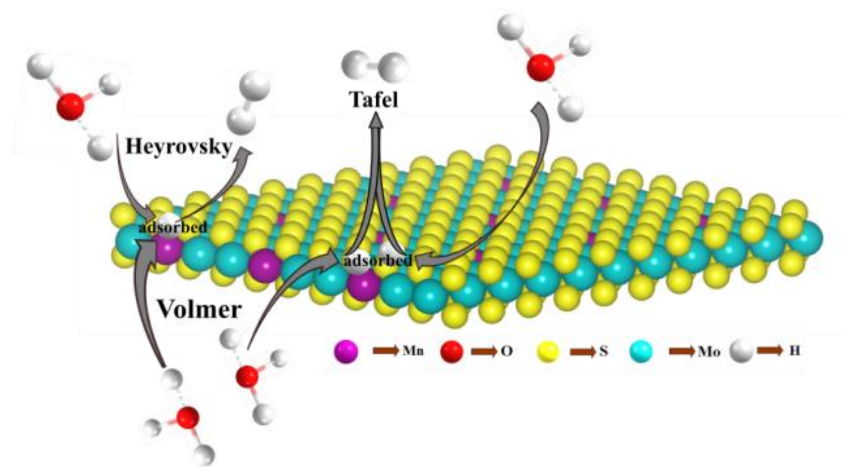